\documentclass{edp-conf}
\usepackage{graphicx}
\begin{document}

\TitreGlobal{SF2A 2001}

\title{The Milky Way as the Ultimate Extragalactic Source} 
\runningtitle{The Extragalactic Milky Way}
\author{Boudewijn F. Roukema}
\address{DARC/LUTH, Observatoire de Paris-Meudon, 
5, place Jules Janssen, F-92195 Meudon Cedex, France}
%
\maketitle
\begin{abstract} 
The GAIA satellite will observe the Galaxy and its closest satellites
in great detail. This should allow (i) dating past events of dwarf
galaxies merging or interacting with the Galaxy, and much improved
orbital parameters of the nine dwarf spheroidals and the Magellanic
Clouds and (ii) dating of active galactic nucleus (AGN) phases of the
Galaxy several Gigayears in the past, by detecting coeval star
formation (e.g. open star clusters) that occurred along 
kpc scale Galactic
jets. Both predictions (i) and (ii) of past history will be highly
valuable for searching for or excluding topologically lensed images of
the Milky Way at high redshift.
\end{abstract}
%
One of the most exciting goals of extragalactic research,
to which the GAIA satellite may provide the key, will be to try 
to detect the Galaxy {\em as an extragalactic object}, due to 
topological lensing by the global geometry of the Universe 
(assuming a perturbed 
Friedmann-Lema\^{\i}tre hot big bang cosmological model).

\begin{figure}
\resizebox{\textwidth}{!}{\includegraphics{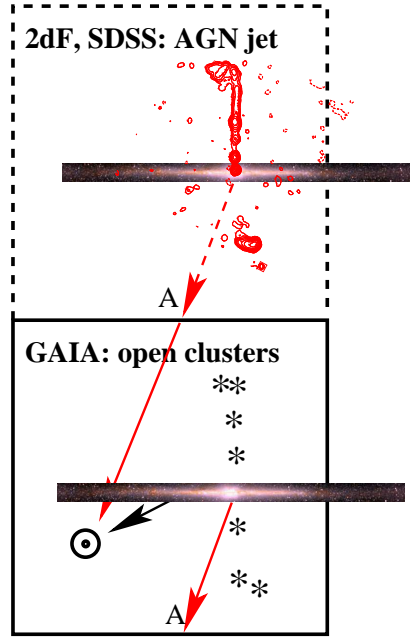}}
\caption[]{For clarity, an exaggeratedly small toroidal universe, about 
16kpc in size, is shown; COBE constraints indicate that the 
real Universe's size 
is about 1$h^{-1}\,$Gpc or greater (Roukema 2000\cite{Rouk00}).
The solid outline (lower square) includes the entire physical
Universe (fundamental polyhedron); the dark arrow shows the geodesic
to the observer at the Sun. The dotted outline (upper square) 
shows a topological 
image of the universe, in apparent space; the gray arrow 
shows the (long time delay) 
geodesic from this ``high'' redshift (early epoch), 
extragalactic, AGN
phase image of the Galaxy, with a double lobe jet $\sim16$kpc in full
length.  In the Galactic image of the Galaxy, 
open star clusters are shown by asterisks following the shape of
the ``high'' redshift jet, though in reality, they would probably 
have been through several orbits since the AGN phase during
which they formed.
}
\label{f-gaia}
\end{figure}

For a recent classification and reference list of
observational methods and preliminary 
results in constraining the global geometry of the Universe, 
see Roukema (2001)\cite{Rouk01}. 
For the specific goal of finding topological
images of our Galaxy, see Wichoski (2000)\cite{Wich00}.

For several decades to come,  
the most systematic ``high'' redshift ($0.5 < z < 3$) 
catalogues of ``extragalactic'' objects 
will remain those of quasars 
Although (i) GAIA's dating of dwarf galaxy merger/interaction 
events will help in the search for extragalactic images of the Galaxy,
(ii) GAIA's dating of AGN phases several Gigayears in the past, e.g.
by detecting open clusters formed along kpc scale 
Galactic jets, will be 
highly valuable for either finding or excluding the possibility 
of an extragalactic quasar image of the Galaxy. 
(The young stars a few million years old at the galactic centre 
would only be useful if the size of the Universe were about a Mpc,
which is clearly not the case!)



\end{document}